\documentclass[showkeys,eqsecnum,showpacs,prd,superscriptaddress]{revtex4}
\usepackage[dvips]{graphicx}
\usepackage{color}
\usepackage{amsmath}
\usepackage[dvips]{graphicx}
\begin{document}

\title{Cosmological Constant as a Free Thermodynamical Variable}

\author{Jarmo M\"akel\"a} 

\email[Electronic address: ]{jarmo.makela@puv.fi}  
\affiliation{Vaasa University of Applied Sciences, Wolffintie 30, 65200 Vaasa, Finland}

\begin{abstract}

We consider a possibility that the cosmological constant may not be a constant, but a free thermodynamical variable. To this end we construct a microscopic model of a spacelike two-sphere just inside of the cosmological horizon of the de Sitter spacetime. In our model the spacelike two-sphere in question is assumed to consist of discrete constituents, each of them contributing to the two-sphere an area, which is an integer times a constant. Using our model we obtain a thermodynamical equation of state for the cosmological constant. Our equation of state implies that the cosmological constant has a certain positive lower bound, which depends on the number of the constituents of the two-sphere.

\end{abstract}


\maketitle

\section{Introduction}

    The cosmological constant has a long history. It was first introduced by Albert Einstein in 1917, when he attempted to create
a static model, based on his recently discovered general theory of relativity, of the universe. \cite{yy} By means of the cosmological 
constant $\Lambda$ one may write Einstein's field equation as:
\begin{equation}
R_{\mu\nu} - \frac{1}{2}g_{\mu\nu} R + \Lambda g_{\mu\nu} = \frac{8\pi G}{c^4}T_{\mu\nu}.
\end{equation} 
Unfortunately, Einstein's expectations for the static  universe did not materialize, and finally the discovery of the expansion of the universe made Einstein to abandon the cosmological constant altogether. For decades it seemed that the cosmological constant must be so small that it has no observational consequences, and the general opinion 
was that the cosmological constant should  be neglected. Nevertheless, there is nothing in the general principles used to justify Einstein's
field equation, which would prevent one from adding the cosmological constant to the field equation, and so the problem was why the cosmological constant should be vanish. Actually, ingenious arguments were expressed to the effect that the cosmological constant must be exactly zero. \cite{kaa}

   Everything changed at the end of the 1990's, when it was discovered, quite unexpectedly, that the universe is not only expanding, but its expansion is {\it accelerating}. \cite{koo} This remarkable observation, which brought to Saul Perlmutter, Brian P. Schmidt and Adam Riess the Nobel Prize in Physics of the year 2011, seemed to suggest that the cosmological constant may not be zero after all, but it may have a certain positive value. Indeed, if one equips Einstein's field equation with a positive cosmological constant, one finds that the expansion of a universe with a constant spatial curvature must be accelerating. Ultimately, the effects of the cosmological constant will overshadow the effects of the matter fields, and spacetime is described, in effect, by the  de Sitter spacetime, which is an empty spacetime with a positive cosmological constant and constant spatial curvature.

   Unfortunately, an inclusion of a positive cosmological constant in Einstein's field equation brings along several problems. One of them is that positive cosmological constant has similar effects as does a matter field with a certain energy density. \cite{nee} So it appears as if the universe contained an enormous amount of energy, which causes an accelerating expansion of the universe. \cite{vii} This energy is known as {\it dark energy}, and it is generally estimated to consist at least 70 per cents of the total energy of the universe.

   In this paper we shall argue that dark energy is not necessarily needed to explain the accelerating expansion of the universe. Instead, the accelerating expansion of the universe might probably be regarded as a quantum effect of gravitation, and it 
arises from the fundamentally discrete structure of space and time. \cite{seite} In short, the idea of this paper is that the cosmological "constant" $\Lambda$ is not really a constant, but a free thermodynamical variable, and if one assumes that spacetime consists of certain fundamental constituents with certain quantum states, the principles of thermodynamics and statistical physics imply that the cosmological constant is  positive and bounded from below. To support this central thesis of ours we consider a spacelike two-sphere just inside of the cosmological horizon of the de Sitter spacetime. For the sake of brevity and simplicity we shall call the two-sphere in question as a {\it shrinked horizon}. We construct the shrinked horizon out of a fixed number of discrete constituents, each of them contributing to the shrinked horizon an area, which is an integer times a constant. In this sense our model of the shrinked horizon of the de Sitter spacetime is similar to the models of the stretched horizons of the Schwarzschild and the Reissner-Nordstr\"om black holes in Refs. \cite{kasi} and \cite{ysi}. Using our model we obtain an explicit, analytic expression for the partition function of the de Sitter spacetime. Our partition function implies, among other things, a thermodynamical equation of state for the cosmological constant. The equation of state, in turn, implies that for a fixed number of constituents of the shrinked horizon the cosmological constant must be positive and bounded from below.  

          Unless otherwise stated, we shall always use the natural units, where $\hbar = c = G = k_B = 1$. 

                  \section{Shrinked Horizon}

                In the static coordinates the line element of the de Sitter spacetime may be written as: \cite{kymmenen}
\begin{equation}
ds^2 = -\left(1 - \frac{\Lambda}{3}r^2\right)\,dt^2 + \frac{dr^2}{1 - \frac{\Lambda}{3}r^2} + r^2\,d\theta^2 
+ r^2\sin^2\theta\,d\phi^2,
\end{equation}
where $\Lambda > 0$ is the cosmological constant, and $\theta$ and $\phi$ are the spherical coordinates.  The de Sitter spacetime has the {\it cosmological horizon}, where
\begin{equation}
r = r_C := \sqrt{\frac{3}{\Lambda}}.
\end{equation}

      Consider now the proper acceleration of an observer with constant coordinates $r$, $\theta$ and $\phi$ in the de Sitter spacetime such that $r < r_C$. The only non-vanishing component of the future pointing unit tangent vector field $u^\mu$ of the congruence of the world lines of observers with constant $r$,  $\theta$ and $\phi$ is 
\begin{equation}
u^t = \left(1 - \frac{\Lambda}{3}r^2\right)^{-1/2},
\end{equation}
and therefore the only non-vanishing component of the proper acceleration vector field 
\begin{equation}
a^\mu := u^\alpha u^\mu_{;\alpha}
\end{equation}
of this congruence is
\begin{equation}
a^r = u^t u^r_{;t} = -\frac{\Lambda}{3}r.
\end{equation}
The vector field $a^\mu$ is spacelike and it has the norm
\begin{equation}
a := \sqrt{a_\mu a^\mu} = \left(1 - \frac{\Lambda}{3}r^2\right)^{-1/2}\frac{\Lambda}{3}r.
\end{equation}
This is the acceleration measured by our observer at rest with respect to the static coordinates for a particle in a radial free fall. The minus sign on the right hand side of Eq. (2.5) indicates that particles in a radial free fall accelerate radially outwards in the de Sitter spacetime.

   Our idea is to consider the cosmological constant $\Lambda$ as  a free variable of our model. In what follows, we shall investigate the spacelike two-spheres with constant $r$ and $t$ just inside of the cosmological horizon, where $r = r_C$. For the sake of simplicity we shall call the two-sphere in question as a {\it shrinked horizon} of the de Sitter spacetime. We shall keep the proper acceleration $a$ given by Eq. (2.6) as a constant on the shrinked horizon. In other words, if the cosmological constant $\Lambda$ is changed, so is the radius $r$ of the shrinked horizon as well,  but in such a way that the proper accelration $a$ remains unchanged. This means that if the cosmological constant $\Lambda$ takes on an infinitesimal change $d\Lambda$, the radius $r$ of the shrinked horizon takes on an infinitesimal change $dr$ such that
\begin{equation}
da = \frac{\partial a}{\partial r}\, dr + \frac{\partial a}{\partial \Lambda}\,d\Lambda = 0,
\end{equation}
and Eqs. (2.2) and (2.6) imply:
\begin{equation}
\frac{dr}{dr_C} = \frac{dr}{d\Lambda}\frac{d\Lambda}{dr_C} = \frac{2r}{r_C}\left(1 - \frac{1}{2}\frac{r^2}{r_C^2}\right).
\end{equation}
So we find that
\begin{equation}
\lim_{r\rightarrow r_C^-}\left(\frac{dr}{dr_C}\right) = 1,
\end{equation}
which means that the shrinked horizon $a = constant$ originally close to the cosmological horizon, where $r = r_C$ will stay close to the cosmological horizon, when the cosmological constant $\Lambda$ is varied. In this sense our shrinked horizon is well chosen.

     In this paper we shall consider the thermodynamics of the de Sitter spacetime from the point of view of an observer at rest on the shrinked horizon $a = constant$. The main reason for this decision of ours is that observers with the same proper acceleration may be considered, in some sense, equivalent, and hence the possible changes in the thermodynamical quantities measured by the observer reflect the changes in the true thermodynamical properties of the de Sitter spacetime, rather than the changes in the state of motion of the observer. We shall see later that our decision brings along a considerable simplification in the calculations, and from the results obtained in the frame of reference, where $a = constant$ one may easily predict the thermodynamical observations, which will be made by observers far from the cosmological horizon.

       \section{Energy}

       The concept of energy plays a central role in the thermodynamical considerations of all systems. Our next aim is to find an expression for the energy of the de Sitter spacetime from the point of view of an observer at rest on its shrinked horizon $a = constant$. 

    We shall take the energy of the de Sitter spacetime from the point of view of our observer to be
\begin{equation}
E = \frac{a}{8\pi}A,
\end{equation}
where $A$ is the area of the shrinked horizon. Our expression is similar to the one obtained in Ref. \cite{kasi} for the energy of the Schwarzschild black hole from the point of view of an observer on its stretched horizon $a = constant$. In Ref. \cite{ysi} the same expression was found for the energy of the Reissner-Nordstr\"om black hole, and in Ref. \cite{yytoo} for the energy of the Kerr-Newman black hole.

  Eq. (3.1) may be justified by means of the {\it Brown-York energy} \cite{kaatoo}
\begin{equation}
E_{BY} := -\frac{1}{8\pi}\oint_S(k - k_0)\,d{\cal A}.
\end{equation}
In Eq. (3.2) we have integrated over a closed, spacelike two-surface $S$ embedded in a spacelike hypersurface, where the time $t$ is a constant. $k$ is the trace of the exterior curvature tensor on the two-surface, and $k_0$ the trace of the exterior curvature tensor on the two-surface, when the two-surface has been embedded in flat spacetime. $d{\cal A}$ is the area element on the two-surface. In stationary spacetimes the Brown-York energy may be understood as the gravitational energy enclosed inside of the two-surface. For a two-sphere $r = constant$ in the de Sitter spacetime the only non-zero components of the exterior curvature tensor are, when $r < r_C$:
\begin{subequations}
\begin{eqnarray}
k_{\theta\theta} &=& -\left(1 - \frac{\Lambda}{3}r^2\right)^{-1/2}\Gamma_{\theta\theta}^r = \left(1 - \frac{\Lambda}{3}r^2\right)^{1/2}r,\\
k_{\phi\phi} &=& -\left(1 - \frac{\Lambda}{3}r^2\right)^{-1/2}\Gamma_{\phi\phi}^r = \left(1 - \frac{\Lambda}{3}r^2\right)^{1/2}r\sin^2\theta,
\end{eqnarray}
\end{subequations}
and therefore the trace of the exterior curvature tensor is:
\begin{equation}
k = k_\theta^{\,\theta} + k_\phi^{\,\phi} = \frac{2}{r}\sqrt{1 - \frac{\Lambda}{3}r^2}.
\end{equation}
When the two-sphere is embedded in flat spacetime, the trace of the exterior curvature tensor is
\begin{equation}
k_0 = \frac{2}{r},
\end{equation}
and because the area of the two-sphere is
\begin{equation}
A = 4\pi r^2,
\end{equation}
the Brown-York energy takes the form:
\begin{equation}
E_{BY} = r\left(1 - \sqrt{1 - \frac{\Lambda}{3}r^2}\right).
\end{equation}

   Consider now how the Brown-York energy will change in the infinitesimal changes of the cosmological constant $\Lambda$. Differentiating the right hand side of Eq. (3.7) with respect to $\Lambda$ one finds that if the cosmological constant takes on an infinitesimal change $d\Lambda$, the Brown-York energy undergoes a change
\begin{equation}
dE_{BY} = \frac{1}{6}\left(1 - \frac{\Lambda}{3}r^2\right)^{-1/2}r^3\,d\Lambda,
\end{equation}
and we may write the Brown-York energy in an integral form:
\begin{equation}
E_{BY} = \int_0^\Lambda \frac{1}{6}\left(1 - \frac{{\tilde \Lambda}}{3}\right)^{-1/2} r^3\,d{\tilde \Lambda}.
\end{equation}
When the cosmological constant $\Lambda$ is considered as a free variable, we may therefore understand the Brown-York energy as the energy downloaded inside of the two-sphere $r = constant$, when the cosmological constant is increased from zero to its given value. Acoording to Eq. (2.9) the change $d\Lambda$ in the cosmological constant implies the change $dr_C$ in the radius $r_C$ of the cosmological horizon such that
\begin{equation}
d\Lambda = \frac{d\Lambda}{dr_C}\,dr_C = -2\frac{\Lambda}{r_C}\,dr_C
\end{equation}
 and hence Eq. (3.8) may also be written, according to Eq. (2.6), as:
\begin{equation}
dE_{BY} = -a\frac{r^2}{r_C}\,dr_C,
\end{equation}
where $a$ is the proper acceleration on the two-sphere. Hence the change in the Brown-York energy may be written in terms of the change $dr$ in the radius of the shrinked horizon $a = constant$ as :
\begin{equation}
dE_ {BY} = -\frac{a}{8\pi}\frac{r}{r_C}\frac{dr_C}{dr}\,dA,
\end{equation}
where $dA = 8\pi r\,dr$ is the change in the area of the shrinked horizon. In the limit, where $r\rightarrow r_C$, we may write Eq. (3.12) as:
\begin{equation}
dE_{BY} = -\frac{a}{8\pi}\,dA.
\end{equation}

   The minus sign on the right hand side of Eq. (3.13) indicates that increase in the area $A$ of the shrinked horizon decreases the amount of energy inside of the shrinked horizon. This means that if we increase the shrinked horizon, energy flows outwards through the shrinked horizon. The amount of energy flown during the process, where the shrinked horizon area has been increased by $dA$ is
\begin{equation}
dE = \frac{a}{8\pi}\,dA,
\end{equation}
and hence it follows that the amount of energy flown through the horizon, when its area is increased from zero to $A$ is
\begin{equation}
E = \frac{a}{8\pi}A.
\end{equation}
An identification of this expression as the energy of the de Sitter spacetime from the point of view of an observer on its shrinked horizon $a = constant$ implies Eq. (3.1). 

    \section{The Partition Function}

   In our model we construct the shrinked horizon $a = constant$ of the de Sitter spacetime out of discrete constituents, each of them contributing to the shrinked horizon an area, which is an integer times a constant. As a consequence, the area of the shrinked horizon takes the form:
\begin{equation}
A = \alpha\ell_{Pl}^2(n_1 + n_2 + n_3 +... + n_N),
\end{equation}
where $n_1, n_2,..., n_N$ are non-negative integers, $\alpha$ is a pure number to be determined later, and
\begin{equation}
\ell_{Pl} := \sqrt{\frac{\hbar G}{c^3}} \approx 1.6 \times 10^{-35} m
\end{equation}
is the Planck length. In Eq. (4.1) $N$, which is assumed to be very large, is the number of the constituents of the shrinked horizon. $N$ is assumed to be fixed. The non-negative integers $n_j$ $(j = 1, 2,..., N)$ are quantum numbers determining the quantum states of the constituents. Constituent $j$ is in vacuum, if $n_j = 0$; otherwise it is in an excited state. Hence our model is similar to the one constructed in Ref. \cite{kasi} for the stretched horizon of the Schwarzschild black hole, and in Ref. \cite{ysi} for the stretched horizon of the Reissner-Nordstr\"om black hole.

   We shall assume that the quantum states of the de Sitter spacetime are somehow encoded into the quantum states of its shrinked horizon. The partition function of the de Sitter spacetime takes the form:
\begin{equation}
Z(\beta) = \sum_n g(E_n)e^{-\beta E_n},
\end{equation}
where the index $n$ labels the possible energies $E_n$ of our system, and $\beta$ is the inverse of its temperature. $g(E_n)$ is the number of the degenerate states associated with the energy $E_n$. Using Eqs. (3.1) and (4.1) we find that the possible energies of the de Sitter spacetime from the point of view of an observer on its shrinked horizon are of the form:
\begin{equation}
E_n = n\alpha\frac{a}{8\pi},
\end{equation}
where
\begin{equation}
n := n_1 + n_2 +...+ n_N.
\end{equation}
As in Ref. \cite{kasi} we take the number $g(E_n)$ of the degenerate states associated with the same energy $E_n$ to be the number of the different combinations of the non-vacuum quantum states of the constituents of the shrinked horizon yielding the same energy $E_n$. As a consequence, $g(E_n)$ equals with the number of ways of expressing a given positive integer $n$ as a sum of at most $N$ positive integers. More precisely, $g(E_n)$ is the number of the ordered strings
\begin{equation}
{\cal S} := (n_1,n_2,...,n_m),
\end{equation}
where $1 \le m \le N$, and $n_j \in \lbrace 1, 2, 3,...\rbrace$ such that 
\begin{equation}
n_1 + n_2 +... + n_m = n.
\end{equation} 
It was found in Ref. \cite{kasi} that the number of such strings is
\begin{equation}
g(E_n) = \sum_{m=1}^{N}\left(\begin{array}{cc}n-1\\m-1\end{array}\right),
\end{equation}
whenever $N < n$. In the special case, where $n = N$, we have
\begin{equation}
g(E_n) = \sum_{m=1}^{n}\left(\begin{array}{cc}n-1\\m-1\end{array}\right) = 2^{n-1}.
\end{equation}
When $n \le N$, Eq. (4.9) always holds, no matter what is $N$.

    We are now prepared to write the partition function for the de Sitter spacetime from the point of view of an observer on its shrinked horizon $a = constant$. As in Ref. \cite{kasi}, the partition function turns out to be of the form:
\begin{equation}
Z(\beta) = Z_1(\beta) + Z_2(\beta),
\end{equation}
where
\begin{subequations}
\begin{eqnarray}
Z_1(\beta) :&=& \frac{1}{2}\sum_{n=1}^N 2^{(1-\beta T_C)n},\\
Z_2(\beta) :&=& \sum_{n=N+1}^\infty\left\lbrack\sum_{k=0}^N\left(\begin{array}{cc}n-1\\k\end{array}\right)2^{-n\beta T_C}\right\rbrack,
\end{eqnarray}
\end{subequations}
and we have defined the {\it characteristic temperature}
\begin{equation}
T_C := \frac{\alpha a}{8\pi\ln 2}.
\end{equation}
The partition function $Z(\beta)$ of Eq. (4.10) may be calculated explicitly, and it takes the form: \cite{kasi}
\begin{equation}
Z(\beta) = \frac{1}{2^{\beta T_C} - 2}\left\lbrack 1 - \left(\frac{1}{2^{\beta T_C} - 1}\right)^{N+1}\right\rbrack,
\end{equation}
when $\beta T_C \ne 1$, and 
\begin{equation}
Z(\beta) = N + 1,
\end{equation}
when $\beta T_C = 1$.

\section{Energy and Entropy}

    The expression obtained in  Eqs. (4.13) and (4.14) for the partition function of the de Sitter spacetime is identical to the partition function obtained in Ref. \cite{kasi} for the Schwarzschild black hole. As a consequence, the expressions for the energy
\begin{equation}
E(\beta) = -\frac{\partial}{\partial\beta}\ln Z(\beta)
\end{equation}
and the entropy
\begin{equation}
S(\beta) = \beta E(\beta) + \ln Z(\beta)
\end{equation}
 of the system from the point of view of an observer on the shrinked horizon $a = constant$ are identical to those of the Schwarzschild black hole. The energy per a constituent
\begin{equation}
{\bar E}(\beta) := \frac{E(\beta)}{N}
\end{equation}
takes, in the leading approximation for large $N$, the form:
\begin{equation}
{\bar E}(\beta) = {\bar E}_1(\beta) + {\bar E}_2(\beta),
\end{equation}
where
\begin{subequations}
\begin{eqnarray}
{\bar E}_1(\beta) :&=& \frac{1}{N}\frac{2^{\beta T_C}}{2^{\beta T_C} - 2}T_C\ln 2,\\
{\bar E}(\beta) :&=& - \frac{2^{\beta T_C}}{(2^{\beta T_C} - 1)^{N+2} - 2^{\beta T_C} + 1}T_C\ln 2.
\end{eqnarray}
\end{subequations}
Eqs. (5.4) and (5.5) imply that
\begin{equation}
\lim_{N\rightarrow\infty}{\bar E}(\beta) = 0,
\end{equation}
whenever $T < T_C$, which means that the constituents of the shrinked horizon are efffectively in vacuum, when the temperature $T$ measured by our observer is less than the characteristic temperature $T_C$. However, if $T > T_C$, the average energy ${\bar E}(\beta)$ per a constituent is non-zero, and Eqs. (5.3) and (5.5b) imply that in the leading approximation for large $N$ we have:
\begin{equation}
E(\beta) = \frac{2^{\beta T_C}}{2^{\beta T_C} - 1}NT_C\ln 2.
\end{equation}
An interesting aspect of this result is that it relates the value of the cosmological constant $\Lambda$ to the absolute temperature $T = 1/\beta$ measured by our observer. Since the shrinked horizon lies very close to the cosmological
 horizon of the de Sitter spacetime, we may effectively regard the areas of these two horizons as equals, and Eqs. (2.2), (3.1), (3.6), (4.12) and (5.7) imply:
\begin{equation}
\Lambda = \frac{12\pi}{N\alpha}\frac{2^{\beta T_C} - 1}{2^{\beta T_C}}.
\end{equation}
So we observe that the cosmological constant $\Lambda$ is, in our model, a temperature-dependent quantity, which depends both on the absolute temperature $T$, and on the number $N$ of the constituents of the shrinked horizon. Eq. (5.8) is the thermodynamical equation of state for the cosmological constant $\Lambda$. We shall consider the cosmological implications of Eq. (5.8) in Section 7. 

    It is interesting that the characteristic temperature $T_C$, which was defined in Eq. (4.12), is proportional to the proper
acceleration $a$ of our observer. If we choose the unspecified numerical constant $\alpha$ such that
\begin{equation}
\alpha = 4\ln 2,
\end{equation}
we find:
\begin{equation}
T = T_U,
\end{equation}
where
\begin{equation}
T_U := \frac{a}{2\pi}
\end{equation}
is the {\it Unruh temperature} measured by our observer. \cite{kaatoo1} We shall see in the next section that the Unruh temperature $T_U$ really is the lowest possible temperature measured by our observer. With  the choice (5.9) for the numerical constant $\alpha$ one may show, as in Ref. \cite{kasi}, that the entropy of the de Sitter spacetime is zero, whenever $T < T_C$. However, if $T > T_C$, one finds that between the area $A$ of the shrinked horizon and the entropy $S$ there is the relationship:
\begin{equation}
S(A) = \frac{1}{4\ln 2}A\ln\left(\frac{2A}{2A - A_{crit}}\right) + N\ln\left(\frac{2A - A_{crit}}{A_{crit}}\right),
\end{equation}
where the {\it critical area}
\begin{equation}
A_{crit} := 8N\ln 2
\end{equation}
is the area of the shrinked horizon in the limit, where $T \rightarrow T_C^+$. In this limit $A \rightarrow A_{crit}^+$, and Eq. (5.12) implies:
\begin{equation}
S(A) = \frac{1}{4}A.
\end{equation}
In other words, the entropy of the deSitter spacetime is one-quarter of the area of its cosmological horizon in the limit, where $T \rightarrow T_C^+$. The result is similar to the one obtained for black holes. \cite{kasi, ysi} Originally, the entropy of the de Sitter spacetime was shown to equal with the one-quarter of the area of its cosmological horizon by Gibbons and Hawking in Ref. \cite{kaatoo2}. 

\section{Phase Transition}

  As a we found in Section 5, the constituents of the shrinked horizon are effecticvely in vacuum, and there is no cosmological horizon, when $T < T_C$. However, when $T > T_C$, Eq. (5.7) implies that the average energy per constituent is given in the large $N$ limit, to a very good approximation, by the formula:
\begin{equation}
{\bar E}(\beta) = \frac{2^{\beta T_C}}{2^{\beta T_C} - 1}T_C\ln 2.
\end{equation}
As one may observe, we have
\begin{equation}
\lim_{T\rightarrow T_C^+}{\bar E}(\beta) = 2T_C\ln 2,
\end{equation}
and so we may conclude that the shrinked horizon undergoes a {\it phase transition} at the characteristic temperature $T_C$. The latent heat per constituent associated with this phase transition is
\begin{equation}
{\bar L} = 2T_C\ln 2.
\end{equation}
Using Eqs. (4.4), (4.5) and (4.12) we find that the average
\begin{equation}
{\bar n}(\beta) := \frac{n_1 + n_2 +...+ n_N}{N}
\end{equation}
of the quantum numbers $n_1, n_2,..., n_N$ is related to ${\bar E}(\beta)$ such that
\begin{equation}
{\bar n}(\beta) = \frac{{\bar E}(\beta)}{T_C\ln 2},
\end{equation}
and Eq. (6.3) implies that
\begin{equation}
{\bar n} = 2
\end{equation}
after the phase transition has been completed. So we find that during the phase transition the constituents of the shrinked horizon jump, in average, from the vacuum to the second excited states. Since the constituents of the shrinked horizon are effectively in vacuum, when $T < T_C$, the characteristic temperature $T_C$, which was found to agree with the Unruh temperature $T_U$, is the lowest possible temperature, which the de Sitter spacetime may have from the point of view of an observer on the shrinked horizon. If the temperature of the environment of the shrinked horizon drops below $T_C$, the constituents  of the shrinked horizon begin to perform transitions from the second excited states to the lower states, and the shrinked horizon begins to radiate with the Unruh temperature $T_U$.

     One of the major advantages of our choice to consider the thermodynamics of the de Sitter spacetime from the point of view of an observer on the shrinked horizon, where the proper acceleration $a = constant$ is that the de Sitter spacetime has, from the point of view of our observer, a well defined, fixed phase transition temperature, which depends on the proper acceleration of the observer only. Such temperature would not exist, if the observer had been chosen otherwise. Moreover, we have seen that the calculations are relatively simple: The expression for the energy $E$ of the de Sitter spacetime in Eq. (3.1) was very simple, and the subsequent calculations leading to the explicit expression in Eq. (4.13) for the partition function $Z(\beta)$ of the de Sitter spacetime were easy to perform.

    Eqs. (2.6) and (5.11) imply that we may write the characteristic temperature $T_C$ as:
\begin{equation}
T_C = B\frac{\Lambda r_C}{6\pi},
\end{equation} 
where 
\begin{equation}
B := \left(1 - \frac{\Lambda}{3}r^2\right)^{-1/2}
\end{equation}
is the blue shift factor. The Tolman relation \cite{kootoo} implies that far from the cosmological horizon, {\it i. e.} when $\frac{r}{r_C} \ll 1$, the temperature of the radiation emitted by the horizon is, when the backscattering effects are neglected:
\begin{equation}
T_\Lambda = \frac{\Lambda r_c}{6\pi} = \frac{\sqrt{3\Lambda}}{6\pi},
\end{equation}
where we have used Eq. (2.2). In the SI units Eq. (6.9) takes the form:
\begin{equation}
T_\Lambda = \frac {\sqrt{3\Lambda}}{6\pi}\frac{\hbar}{k_B}.
\end{equation}
With the currently accepted estimate 
\begin{equation}
\Lambda \sim 10^{-35}s^{-2}
\end{equation}
for the cosmological constant one finds:
\begin{equation}
T_\Lambda \sim 10^{-30}K,
\end{equation}
which is very low, indeed.

\section{Equation of State}

   Consider now in details the thermodynamical equation of state found in Eq. (5.8) for the cosmological constant. In the SI units, using Eq. (5.9), we may write Eq. (5.8) as:
\begin{equation}
\Lambda = \frac{3\pi c^2}{N\ell_{Pl}^2\ln 2}\frac{2^{\beta k_BT_C} - 1}{2^{\beta k_BT_C}},
\end{equation}
where $\ell_{Pl}$ is the Planck length, which was defined in Eq. (4.2). Eq. (6.13) implies:
\begin{equation}
\lim_{T\rightarrow T_C^+}\Lambda = \frac{3\pi c^2}{2N\ell_{Pl}^2\ln 2},
\end{equation}
and therefore we are able to write the cosmological constant as a function of the number $N$ of the constituents of the shrinked horizon only in the special case, where the temperateration ure measured by our observer on the shrinked horizon agrees with his Unruh temperature. Since $N$ is assumed to be very large, the cosmological constant $\Lambda$ is very small. Even though $N$ is very large, however, $N$ is finite, and therefore the cosmological constant must be non-zero. Hence we may say that in our model non-zero cosmological constant appears as  a natural consequence of the discrete structure of spacetime.

   It is somewhat uncertain what we should regard as the temperature of the cosmological horizon from the point of view of an observer on the shrinked horizon. The majority of physicists would probably hold the view that the temperature in question is simply the Unruh temperature measured by the observer. In that case the constituents of the horizon lie, in average, on the second excited states, and Eqs. (4.1) and (5.9) imply that the average area of an individual constituent is
\begin{equation}
{\bar A} = 8\ell_{Pl}^2\ln 2 \approx 1.4 \times 10^{-69}m^2.
\end{equation}
Using Eq. (7.2) we may write the constituent number $N$ as a function of the cosmological constant $\Lambda$:
\begin{equation}
N = \frac{3\pi c^2}{2\Lambda \ell_{Pl}^2\ln 2},
\end{equation}
and putting $\Lambda \sim 10^{-35}s^{-2}$ we find:
\begin{equation}
N  \sim 10^{122}.
\end{equation}
This is a huge number, but nevertheless it is finite. The reason why the observed value of the cosmological constant is around $10^{-35}s^{-2}$ is, in our model, that the number of the constituents of the cosmological horizon is around
$10^{122}$.

\section{Consistency of the Model}

   As we found in Section 6, the cosmological horizon emits radiation with the characteristic temperature $T_\Lambda$
given in Eq. (6.10). As a consequence, the cosmological constant will increase in time. However, at the beginning of our analysis in Section 2 we assumed that the cosmological constant is strictly constant. We must therefore check, whether 
the increase of the cosmological constant in time as a consequence of the radiation of the cosmological horizon is slow enough such that we may regard the cosmological "constant", in effect, as a true constant. 

   According to the first law of thermodynamics the amount of energy emitted by the cosmological horizon is, from the point of view of an observer far from the horizon:
\begin{equation}
dE_\Lambda = -T_\Lambda\,dS,
\end{equation}
where $dS$ is the change in the entropy $S$ of the horizon. Using Es: (5.14) and (6.10) we find that, in the SI units:
\begin{equation}
dE_\Lambda = -\frac{\sqrt{3\Lambda}}{24\pi}\frac{c^3}{G}\,dA,
\end{equation}
where $dA$ is the change in the area of the cosmological horizon. Writing $dA = 8\pi r_C\,dr_C$, and using Eq. (2.2) we find that the energy emitted in a unit time is:
\begin{equation}
\frac{dE_\Lambda}{dt} = \frac{c^5}{2G}\sqrt{\frac{3}{\Lambda^3}}\frac{d\Lambda}{dt}.
\end{equation}
On the other hand, the energy emitted by the cosmological horizon in a unit time is given by the Stefan-Boltzmann law:
\begin{equation}
\frac{dE_\Lambda}{dt} = \sigma AT^4_\Lambda,
\end{equation}
where $\sigma$ is the Stefan-Boltzmann constant. Comparing Eqs. (8.3) and (8.4) we get:
\begin{equation}
\frac{1}{\Lambda}\frac{d\Lambda}{dt} = \frac{G\hbar^4}{18\pi^3k_B^4c^3}\sigma\sqrt{3\Lambda^3}.
\end{equation}
Again, putting $\Lambda \sim 10^{-35}s^{-2}$ we find:
\begin{equation}
\frac{1}{\Lambda}\frac{d\Lambda}{dt} \sim 10^{-143} s^{-1},
\end{equation}
which means that the time needed for the doubling of the cosmological constant would be around $10^{143}$ seconds. This is a huge amount of time (the present age of the universe is about $10^{17}$ seconds), and hence we may regard the cosmological constant, from the practical point of view, as a true constant of nature. This means that our analysis is self-consistent.

       In the considerations performed so far we have identified the temperature measured by an observer on the shrinked horizon with his Unruh temperature. It is interesting to consider what would happen, if we assumed the shrinked horizon to be in thermal equilibrium with the cosmic microwave background. In that case the temperature measured by our observer for the shrinked horizon would equal with his measurement for the temperature of the cosmic microwave background.   Since the current temperature $T_R \approx 3K$ measured by an observer far from the cosmological horizon for the cosmic microwave background is very much higher than the temperature $T_\Lambda$ in Eq. (6.12), the temperature $T$ measured by the  observer on the shrinked horizon for the cosmic microwave background must be much higher than $T_C$. Since
\begin{equation}
2^x \approx 1 + x\ln 2,
\end{equation}
when $x\ll 1$, we may write Eq. (7.1) as:
\begin{equation}
\Lambda = \frac{3\pi c^2}{N\ell_{Pl}^2}\frac{T_C}{T},
\end{equation}
when $T \gg T_C$. The Tolman relation implies that
\begin{subequations}
\begin{eqnarray}
T_C &=& BT_\Lambda,\\
T &=& BT_R, 
\end{eqnarray}
\end{subequations}
where $B$ is the blue shift factor, defined in Eq. (6.8), for the observer on the shrinked horizon. Hence we find:
\begin{equation}
\Lambda = \frac{3\pi c^2}{N\ell_{Pl}^2}\frac{T_\Lambda}{T_R},
\end{equation}
and Eq. (6.10) implies for the cosmological constant an equation:
\begin{equation}
\Lambda = \frac{\sqrt{3\Lambda} c^5}{2NGk_BT_R},
\end{equation}
which has the solution:
\begin{equation}
\Lambda = \frac{3c^{10}}{4G^2}\frac{1}{(Nk_BT_R)^2}.
\end{equation}
From this equation we may solve $N$ in terms of the cosmological constant $\Lambda$:
\begin{equation}
N = \frac{c^5}{2Gk_BT_R}\sqrt{\frac{3}{\Lambda}}.
\end{equation}
Putting $\Lambda \sim 10^{-35}s^{-2}$ and $T_R = 3K$  we find
\begin{equation}
N \sim 10^{92}.
\end{equation}
This number is 30 orders of magnitude less than the estimate obtained in Eq. (7.5). As a consequence, the average diameter of an individual constituent of the cosmological horizon is about $10^{15}$ times the Planck length, or $10^{-20}m$, which is still much less than the effective size of any known elementary particle. Substituting Eq. (8.14) in Eq. (8.12) we may express Eq. (8.12) numerically as:
\begin{equation}
\Lambda \sim 10^{-35}\frac{1}{T_R^2}K^2 s^{-2}.
\end{equation}

     It is very interesting that according to Eq. (8.15) the cosmological constant is inversely proportional to the square of the absolute temperature $T_R$ of the cosmic microwave background. So we find that when the universe cools down beacuse of its expansion and $T_R$ decreases, the cosmological constant increases, and it seems that the cosmological constant was less in the past than it is now. Unfortunately, such a conclusion would lead to inconsistencies in our model, which may be seen, if we consider the energy density of the cosmic microwave background. Since the energy density of thermal radiation is proportional to the fourth power of its absolute temperature, conservation of energy implies:
\begin{equation}
R^3T_R^4 = constant,
\end{equation}
where $R$ is the scale factor of the universe. Differentiating the both sides of Eq. (8.16) we find:
\begin{equation}
3\frac{dR}{R} = -4\frac{dT_R}{T_R},
\end{equation}
and Eq. (8.12) implies an equation:
\begin{equation}
\frac{d\Lambda}{dR} = \frac{3}{2}\frac{\Lambda}{R},
\end{equation}
which has the solution:
\begin{equation}
\Lambda = CR^{3/2},
\end{equation}
where $C$ is a positive constant. So we observe that the cosmological constant increases surprsingly rapidly as a function of the scale factor $R$. Actually, the increase of the cosmological constant is so rapid that it takes us into a contradiction with our basic assumption, which was that the cosmological constant may be regarded, at least effectively, as a constant. A careful re-analysis, where the possible increase of the cosmological constant in time is taken into account right from the beginning, is therefore needed. \cite{viitoo} Nevertheless, our results are very suggestive, and it would be very nice, if observational cosmologists managed to find evidence for an increasing cosmological constant.

    It may be pretty surprising that as a result of the Hawking-type radiation emitted by the cosmological horizon the cosmological constant $\Lambda$ {\it increases} in time leading to the shrinking of the cosmological horizon. If the entropy of the de Sitter spacetime equals to one-quarter of the area of its horizon and the horizon shrinks, then what will happen to the second law of thermodynamics, which states that the total entropy of an isolated system may never decrease in time?

      The radiation emitted by the cosmological horizon does not violate the second law of thermodynamics any more than does the Hawking radiation of a black hole. The decrease of the entropy of the horizon as a result of the decrease in its area is always less or the same as the increase in the entropy of the radiation. As a consequence, the sum of the entropies of the horizon and the radiation may never decrease in time, and the second law of thermodynamics remains to be valid.

   It should be noted that in our very simple model the universe is not assumed to consist matter at all, but just a positive cosmological constant. In the real universe, of course, there is matter as well, and the area of the cosmological horizon is not determined by the cosmological constant alone, but also by the matter fields. During the course of time matter escapes beyond the cosmological horizon and, as a consequence, the area of the cosmological horizon increases in time in the same way as does the event horizon area of a black hole, when the hole swallows matter. Such processes have been investigated, among other things, in a  recent paper by Mimoso and Pav\'on, and it was found that the entropy of the horizon plus that of radiation and matter inside it increases and is concave. \cite{kuutoo}

 \section{Concluding Remarks}

   In this paper we have considered the thermodynamics of the de Sitter spacetime from the point of view of an observer on a spacelike two-sphere, which is just inside of the cosmological horizon. For the sake of brevity and simplicity we called the two-sphere on question as the {\it shrinked horizon} of the de Sitter spacetime. Our idea was to consider the cosmological constant $\Lambda$ as a free thermodynamical variable of the system, and we assumed that when the cosmological constant varies, the radius of the shrinked horizon will also change, but in such a way that the proper accceleration $a$ of an observer on the shrinked horizon stays unchanged. We constructed the shrinked horizon out of discrete constituents, each of them contributing to the shrinked horizon an area, which is an integer times a constant. In this sense our model was similar to those constructed previously for the Schwarzschild \cite{kasi} and the Reissner-Nordstr\"om \cite{ysi} black holes. Using our model we managed to obtain an explicit, analytic expression for the partition function of the de Sitter spacetime. 

  Our partition function implied, among other things, that an observer on the shrinked horizon measures for the cosmological horizon a certain minimum temperature, which is proportional to the observer's proper acceleration $a$, and may be identified as the observer's Unruh temperature. When the temperature of the horizon equals, from the point of view of our observer, with the observer's Unruh temperature, the entropy of the de Sitter spacetime is, in the natural units, exactly one-quarter of the area of the cosmological horizon.

    The most important result obtained in this paper concerned the properties of the cosmological constant. We found that
if the temperature measured by an observer on the shrinked horizon equals with his Unruh temperature, the cosmological constant is inversely proportional to the number of the constituents of the shrinked horizon. However, because the number of the constituents of the shrinked horizon was assumed to be finite -even though very large- it follows that the cosmological constant must necessarily be non-zero and bounded from below. At the Unruh temperature the constituents of the shrinked horizon are Planck-size objects, and using the currently accepted estimate $10^{-35}s^{-2}$ for the cosmological constant we obtained for the number of the constituents of the shrinked horizon which, for all practical purposes, may be identified with the cosmological horizon, an estimate $10^{122}$.

   Taken as a whole, our model seems to suggest that dark energy is not necessarily needed to explain the acceleration of the expansion of the universe, but the accelarating expansion of the universe may be a simple consequence of the fundamentally discrete nature of spacetime, and of  the principles of thermodynamics. According to this view spacetime has certain fundamental constituents -sort of "atoms of spacetime"- with certain quantum  states, and if one considers the cosmological "constant" as a free thermodynamical variable, the principles of thermodynamics imply that $\Lambda$ is positive and bounded from below. In this sense we may regard the accelerating expansion of the universe as a quantum effect of gravitation.


\begin{thebibliography}{20}
\bibitem{seite} The idea that the accelerating expansion of the universe may be caused by a discrete structure of spacetime is not new. It has been advocated, among others, by T. Padmanabhan, arXiv:1206.4916, arXiv:1207.0505, arXiv:1208.1375, arXiv:1210.4174.
\end{thebibliography}
\end{document}